\documentclass{aa} 
\usepackage{graphicx}
\usepackage{natbib}  
\usepackage[varg]{txfonts}
\usepackage[breaklinks]{hyperref}

\def\Mgd  {\ion{Mg}{ii}}

\def\Cad  {\ion{Ca}{ii}}

\def\Feu  {\ion{Fe}{i}}

\def\Yd   {\ion{Y}{ii}}
\def\Zrd  {\ion{Zr}{ii}}

\def\Teff  {$T_\mathrm{eff}$}

\def\loggf {$\log gf$}
\def\vt    {$\rm v_{t}$}
\def\kms   {$\rm km\,s^{-1}$}

\begin{document}

\title{Neutron-capture elements in a sample of field metal-poor N-rich dwarfs.
\thanks{Based on observations collected at the European Organisation for Astronomical Research in the Southern Hemisphere (Archives of programmes 090.B-0504(A) PI Chaname; 095.D-0504(A) PI Melendez; 076.B-0166(A) PI Pasquini; 067.D-0086(A) PI Gehren; 071.B-0529(A) PI Silva; 065.L-0507(A) PI Primas), and collected at the W. M. Keck Observatory, archive programme G401H, PI Melendez. One star was also observed at ESO with the spectrograph ESPRESSO, programme 107.22RU.001 PI Spite, and two stars were observed at the Observatoire de Haute Provence (Archives of programme 10A.PNPS.HALB, PI Halbwachs) and at the Narval spectrograph of the Observatoire du Pic du Midi, programme L172N04, PI Spite.}
}

\author{
M. Spite \inst{1}\and 
P. Bonifacio \inst{1}
E. Caffau \inst{1}\and
P. Fran\c cois  \inst{2}
 }
\institute {
GEPI, Observatoire de Paris, PSL Research University, CNRS,
Place Jules Janssen, 92190 Meudon, France
\and
GEPI, Observatoire de Paris, PSL Research University, CNRS, 
Univ. Paris Diderot, Sorbonne Paris Cité, 61 avenue de l'Observatoire, 75014, Paris, France
}

\authorrunning{Spite M. et al.}
\titlerunning{Heavy elements in Metal-poor N-rich dwarf stars}

\abstract
{The aim of this work is to measure the abundances of n-capture elements in a sample of six metal-poor N-rich dwarfs that were formed in globular clusters, and  subsequently became unbound from the cluster. These N-rich stars, HD\,25329, HD\,74000, HD\,160617, G\,24-3, G\,53-41, and G\,90-3, were previously studied in  Paper I.}
 {The abundances of the n-capture elements in these stars were compared to the abundances in normal metal-poor dwarfs and in globular cluster stars in the same metallicity range in order to find  evidence of an enrichment of the material from which  these N-rich stars were formed, by the ejecta of massive asymptotic giant branch stars (AGB) inside the cluster.}
{The abundances of 15 elements,  from Sr to Yb,  were derived line by line by comparing the observed profiles to synthetic spectra in a sample of six metal-poor N-rich dwarfs and nine classical metal-poor dwarfs.}  
{We show that, generally speaking, the behaviours of the intermediate metal-poor stars here studied and the extremely metal-poor stars are very different. In particular, the scatter of the [X/Fe] ratios is much smaller since many more stars contribute to the enrichment.\\
Among our six metal-poor N-rich stars, three stars (G24-3 and HD\,74000 and maybe also HD\,160617) present an enrichment in elements formed by the s-process, typical of a contribution of AGB stars, unexpected at the metallicity of these stars. This suggests that the intracluster medium from which these stars were formed was enriched by a first generation of massive AGB stars. \\
Another N-rich star, G53-41, is also rich in s-process elements, but since it is more metal-rich this could be due to the normal galactic enrichment by low-mass AGB stars before the formation of the cluster. 
In contrast, two stars (G\,90-3 and HD\,25329) have an abundance pattern compatible with a pure r-process such as that seen in   metal-poor stars with [Fe/H]<--1.5.}
{}

\maketitle
%
\section{Introduction}
In a previous paper \citep{SpiteSC22} (hereafter  Paper I) we  showed that, in the field of our Galaxy,  nitrogen-rich dwarfs that are not enriched in carbon  present the same characteristics as the second-generation stars in  globular clusters (GCs). 
In these stars C and O are slightly deficient, but the scatter of the relation [(C+N+O)/Fe] versus [Fe/H]\footnote{For each element X, we adopted the classical notations:~~ A(X)=(log (N(X)/N(H)) + 12),~~~~$\rm [X/H]=A(X)_{\star}-A(X)_{\odot}$ ~~and~~~ [X/Fe]=[X/H]--[Fe/H].} is very small, as  is expected if the N-enrichment is the result of  pollution by a CNO processed material.
These stars show an excess of Na and sometimes of Al; a Na-O anticorrelation is observed similar to the anticorrelation observed in NGC6752.\\

In this second paper we explore the behaviour of the n-capture elements from Sr to Yb in the same six nitrogen-rich dwarfs in order to compare them to the classical metal-poor dwarfs and to the  second generation of GC stars. Eight GCs were used for this comparison.  Unfortunately, generally speaking, for a given GC star we do not know  its abundance in heavy elements and its N abundance at the same time, and this limits the comparability.

The n-capture elements are mainly built by the  s-process (slow compared to the $\beta$ decay of the affected nucleus) or by the r-process (rapid process).   
The r-process occurs on a very short timescale in violent events, explosions of massive supernovae, and mergers (e.g.  two neutron stars or  a neutron star with a black hole or gamma ray burst) \citep{CowanSL21,ArconesThielemann23}.\\ 
On the contrary, the low- and intermediate-mass  (LIM) stars
on their asymptotic giant branch  (AGB) phase seem to be the main site of the s-process \citep{ArnouldGoriely20,PrantzosAC20,ArconesThielemann23}. Moreover a weak s-process in rotating massive very metal-poor stars \citep{FrischknechtHP16} could explain the abundance of the first peak n-capture elements in very metal-poor stars \citep[see e.g. ][]{SpiteSB18}.

The abundances of the n-capture elements have been studied in several GCs, in particular
  M2 \citep{YongRG14}, M15 \citep{SobeckKS11,WorleyHS13},  M19 \citep{JohnsonRP15}, M22 \citep{RoedererMS11}, M92 \citep{RoedererSneden11}, NGC\,104 (47 Tuc) \citep{JamesFB04b}, NGC\,4833 \citep{RoedererThompson15}, NGC\,5286 \citep{MarinoMK15}, NGC\,5824 \citep{RoedererMB16}, NGC\,6397 \citep{JamesFB04b},  NGC\,6266 (M62) \citep{YongABDC14}, NGC\,6752 \citep{JamesFB04a}, and  NGC\,7089 \citep{YongRG14}.\\
In  clusters with no metallicity dispersion, generally the abundances of the n-capture elements do not show strong variations from star to star; however, there are exceptions, for example  NGC\,5824 \citep{RoedererMB16}.
\\ 
This abundance pattern in GCs is generally explained by the combination of a first production by some form of r-process, sometimes enriched later with s-process material produced and ejected by relatively massive AGB stars inside the cluster \citep[see e.g.][]{RoedererMB16}.

If we assume that the N-rich stars here studied were formed within a globular cluster (see Paper I), the aim of the present paper is to look for traces of n-capture element enrichment by massive AGB stars inside the cluster from which these stars originate. 

\begin{table}
\caption[]{ Atmospheric parameters of the dwarf stars studied in this paper.}
\label{Tab:AtmosParam}
\centering
\begin{tabular}{lccclcrrrrrrr}
\hline
 Ident.&       Teff  & log g   & \vt  &[Fe/H]\\
       &             &         & \kms &      \\        
\hline
 {\it N-rich dwarfs}\\
 HD\,25329   & 4870 & 4.75 & 0.8 & -1.72\\
 HD\,74000   & 6260 & 4.30 & 1.2 & -2.00\\
 HD\,160617  & 6000 & 3.90 & 1.2 & -1.80\\
 G\,24-3     & 6030 & 4.40 & 0.9 & -1.58\\
 G\,53-41    & 6050 & 4.40 & 0.9 & -1.19\\
 G\,90-3     & 5900 & 3.70 & 1.2 & -2.18\\
\hline 
{\it Normal dwarfs}\\ 
 HD\,19445   & 6070 & 4.40 & 1.3 & -2.15\\
 HD\,76932   & 6000 & 4.10 & 1.3 & -0.94\\
 HD\,84937   & 6300 & 4.00 & 1.3 & -2.25\\
 HD\,94028   & 6050 & 4.30 & 1.2 & -1.40\\
 HD\,97916   & 6400 & 4.35 & 1.2 & -0.75\\
 HD\,108177  & 6275 & 4.30 & 1.1 & -1.65\\
 HD\,140283  & 5750 & 3.70 & 1.4 & -2.57\\
 HD\,166913  & 6230 & 4.20 & 1.3 & -1.50\\
 HD\,218502  & 6300 & 4.10 & 1.3 & -1.85\\
\hline
\end{tabular}
\end{table}

\begin{table*}
\caption[]{Abundances of the elements:  A(X)= 12 + log (N(X)/N(H)),  [X/H]=$A(X)_{*}-A(X)_{\odot}$, and [X/Fe]=[X/H]--[Fe/H]. The six N-rich dwarfs are at the top of the table, followed by the reference stars with a normal N abundance. The adopted solar abundances are in boldface}
\label{Tab:abund}
\centering
\begin{tabular}{l@{~~}c@{~~~~~~~}c@{~}r@{~~~~~}c@{~}r@{~~~~~}c@{~}r@{~~~~~}c@{~}r@{~~~~~}c@{~}r@{~~~~~}c@{~}r@{~~~~~}c@{~}c@{~~~~~}c@{~}c@{~~~~~}c@{~}c@{~~~~~}c@{~}c@{~~~~~}c@{~}c@{~~~~~~}c@{~}}

\hline
 Atomic number&        &   38    &         & 39    &        &  40    &          &  42    &        &  44   &        \\
             &[Fe/H]   &   A(Sr) &[Sr/Fe]  &  A(Y) & [Y/Fe] &  A(Zr) &[Zr/Fe]   &  A(Mo) &[Mo/Fe] &  A(Ru) &[Ru/Fe]\\
\hline
  {\bf Sun}    &         & {\bf2.92}&         &{\bf2.21}&      &{\bf2.58} &        &{\bf1.92} &        &{\bf1.84} &\\
{\it N-rich dwarfs}\\
  HD 25329    &  -1.72  &    1.53 &   0.33  &  0.70 &   0.21 &   1.48 &   0.62 &   --   &   --   &   0.83 &   0.71 \\
  HD 74000    &  -2.00  &    1.41 &   0.49  &  0.27 &   0.06 &   1.05 &   0.47 &   0.62 &   0.70 &   0.60 &   0.76 \\
 HD 160617    &  -1.80  &    1.34 &   0.22  &  0.33 &  -0.08 &   1.07 &   0.29 &   0.75 &   0.63 &   0.71 &   0.67 \\
     G24-3    &  -1.58  &    1.51 &   0.17  &  0.61 &  -0.02 &   1.29 &   0.29 &   0.94 &   0.60 &   0.82 &   0.56 \\
    G53-41    &  -1.19  &    1.94 &   0.21  &  1.17 &   0.15 &   1.90 &   0.51 &   1.42 &   0.69 &   1.42 &   0.77 \\
     G90-3    &  -2.18  &    0.97 &   0.23  & -0.17 &  -0.20 &   0.65 &   0.25 &   0.28 &   0.54 &   0.47 &   0.81 \\
{\it Normal dwarfs}\\
  HD 19445    &  -2.15  &    1.07 &   0.39  &  0.12 &   0.06 &   0.82 &   0.38 &   0.47 &   0.70 &   0.50 &   0.81 \\
  HD 76932    &  -0.94  &    2.20 &   0.30  &  1.29 &  -0.02 &   1.97 &   0.28 &   1.56 &   0.58 &   1.43 &   0.53 \\
  HD 84937    &  -2.25  &    0.97 &   0.39  & -0.05 &  -0.01 &   0.65 &   0.31 &   0.25 &   0.58 &   0.40 &   0.81 \\
  HD 94028    &  -1.40  &    1.79 &   0.36  &  0.90 &   0.09 &   1.56 &   0.37 &   1.26 &   0.74 &   1.03 &   0.59 \\
  HD 97916    &  -0.75  &    2.24 &   0.07  &  1.37 &  -0.09 &   1.87 &   0.04 &    --  &    --  &    --  &    --  \\
 HD 108177    &  -1.65  &    1.61 &   0.34  &  0.57 &   0.01 &   1.33 &   0.40 &   1.09 &   0.82 &   0.80 &   0.61 \\
 HD 140283    &  -2.57  &    0.08 &  -0.18  & -0.78 &  -0.42 &  -0.01 &  -0.07 &  -0.58 &   0.07 &    --  &    --  \\
 HD 166913    &  -1.50  &    1.80 &   0.38  &  0.80 &   0.09 &   1.48 &   0.40 &   1.16 &   0.74 &   1.06 &   0.72 \\
 HD 218502    &  -1.85  &    1.42 &   0.35  &  0.31 &  -0.05 &   1.07 &   0.34 &   0.86 &   0.79 &   0.61 &   0.62 \\
\\
\hline
Atomic number&          &  46    &        &  47    &        &  56    &        &  57    &        &  58    &        \\
             &[Fe/H]    &  A(Pd) &[Pd/Fe] &  A(Ag) &[Ag/Fe] &  A(Ba) &[Ba/Fe] &  A(La) &[La/Fe] &  A(Ce) &[Ce/Fe] \\
\hline
  {\bf Sun}    &          &{\bf1.66}&        &{\bf0.94}&        &{\bf2.17}&        &{\bf1.14}&        &{\bf1.60}  \\
{\it N-rich dwarfs}\\
 HD 25329    &  -1.72   &   0.37 &   0.43 & --     &   --   &   0.81 &   0.36 &  -0.11 &   0.47 &  0.54: &   0.66:\\
 HD 74000    &  -2.00   &  --    &    --  &  -0.19 &   0.87 &   0.32 &   0.15 &  -0.53 &   0.33 &  -0.10 &   0.30 \\
HD 160617    &  -1.80   &   0.04 &   0.18 &  -0.40 &   0.46 &   0.67 &   0.30 &  -0.28 &   0.38 &   0.18 &   0.38 \\
    G24-3    &  -1.58   &   0.01 &  -0.07 &   --   &   --   &   0.79 &   0.20 &  -0.22 &   0.22 &   0.20 &   0.18 \\
   G53-41    &  -1.19   &   0.81 &   0.34 &   0.32 &   0.57 &   1.65 &   0.67 &   0.52 &   0.57 &   0.87 &   0.46 \\
    G90-3    &  -2.18   &  -0.51 &   0.01 &    --  &   --   &  -0.09 &  -0.08 &  -0.95 &   0.09 &  -0.50 &   0.08 \\
{\it Normal dwarfs}\\
  HD 19445   &   -2.15  &   -0.54&   -0.05&    --  &      --&    0.00&   -0.10&   -0.88&    0.13&   -0.30&    0.25 \\
  HD 76932   &   -0.94  &    0.86&    0.14&    0.43&    0.43&    1.42&    0.10&    0.46&    0.26&    0.80&    0.14 \\
  HD 84937   &   -2.25  &   -0.39&    0.20&     -- &      --&   -0.22&   -0.22&   -0.74&    0.37&   -0.20&    0.45 \\
  HD 94028   &   -1.40  &    0.32&    0.06&   -0.14&    0.32&    1.02&    0.17&    0.02&    0.28&    0.46&    0.26 \\
  HD 97916   &   -0.75  &     -- &     -- &     -- &      --&    1.57&    0.15&    0.46&    0.07&    0.88&    0.03 \\
 HD 108177   &   -1.65  &    0.22&    0.21&   -0.11&    0.60&    0.53&    0.01&   -0.31&    0.20&    0.19&    0.24 \\
 HD 140283   &   -2.57  &    --  &    --  &     -- &      --&   -1.09&   -0.81&   -1.81&   -0.38&    --  &     --  \\
 HD 166913   &   -1.50  &    0.54&    0.38&   -0.10&    0.46&    0.79&    0.12&   -0.21&    0.15&    0.24&    0.14 \\
 HD 218502   &   -1.85  &   -0.03&    0.16&    --  &     -- &    0.22&   -0.10&   -0.57&    0.14&   -0.18&    0.07 \\
\\
\hline
Atomic number&        &    60    &        &  63   &         &  66    &        &  68    &        &  70    &        \\
            & [Fe/H]  &    A(Nd) &[Nd/Fe] &  A(Eu)& [Eu/Fe] &  A(Dy) &[Dy/Fe] &  A(Er) &[Er/Fe] &  A(Yb) &[Yb/Fe] \\
\hline
  {\bf Sun}  &         & {\bf1.45}&        &{\bf 0.52}&         &{\bf1.14}&        &{\bf0.96}&        &{\bf0.86}& \\
{\it N-rich dwarfs}\\
 HD 25329   &   -1.72 &     0.30 &   0.57 & -0.35:&    0.85:&  -0.25 &   0.33 &  -0.06 &   0.70 &  -0.34 &   0.52 \\
 HD 74000   &   -2.00 &    -0.07 &   0.48 &  -1.30&    0.18 &  -0.93 &  -0.07 &  -0.90 &   0.14 &  -0.87 &   0.27 \\
HD 160617   &   -1.80 &    -0.03 &   0.32 &  -0.70&    0.58 &  -0.11 &   0.55 &  -0.36 &   0.48 &  -0.34 &   0.60 \\
    G24-3   &   -1.58 &     0.22 &   0.35 &  -0.77&    0.29 &  -0.13 &   0.31 &  -0.43 &   0.19 &  -0.37 &   0.35 \\
   G53-41   &   -1.19 &     0.65 &   0.39 &  -0.18&    0.49 &   0.55 &   0.60 &   0.22 &   0.45 &   0.60 &   0.93 \\
    G90-3   &   -2.18 &    -0.56 &   0.17 &  -1.35&    0.31 &  -0.72 &   0.32 &  -0.99 &   0.23 &  -1.22 &   0.10 \\
{\it Normal dwarfs}\\
  HD 19445  &    -2.15&     -0.25&    0.45&  -1.26 &   0.37&   -0.67&    0.34&   -0.65&    0.54&   -1.02&    0.27 \\
  HD 76932  &    -0.94&      0.87&    0.36&   0.00 &   0.36&    0.63&    0.43&    0.49&    0.47&    0.77&    0.85 \\
  HD 84937  &    -2.25&     -0.30&    0.50&  -1.35 &   0.38&   -0.67&    0.44&   -0.79&    0.50&   -0.98&    0.41 \\
  HD 94028  &    -1.40&      0.30&    0.25&  -0.63 &   0.25&   -0.08&    0.18&   -0.22&    0.22&   -0.27&    0.27 \\
  HD 97916  &    -0.75&      0.90&    0.20&  -0.18 &   0.05&    0.59&    0.20&    0.25&    0.04&    0.57&    0.46 \\
 HD 108177  &    -1.65&      0.09&    0.29&  -0.81 &   0.32&   -0.21&    0.30&   -0.40&    0.29&   -0.46&    0.33 \\
 HD 140283  &    -2.57&       -- &     -- &  -2.27 &  -0.28&   -2.20&   -0.77&   -2.01&   -0.40&   -2.27&   -0.56 \\
 HD 166913  &    -1.50&      0.09&    0.14&  -0.64 &   0.34&   -0.06&    0.30&   -0.26&    0.28&   -0.27&    0.37 \\
 HD 218502  &    -1.85&      0.18&    0.58&  -0.99 &   0.34&   -0.37&    0.34&   -0.66&    0.23&   -0.70&    0.29 \\
\hline
\end{tabular}
\end{table*}

\begin{figure}
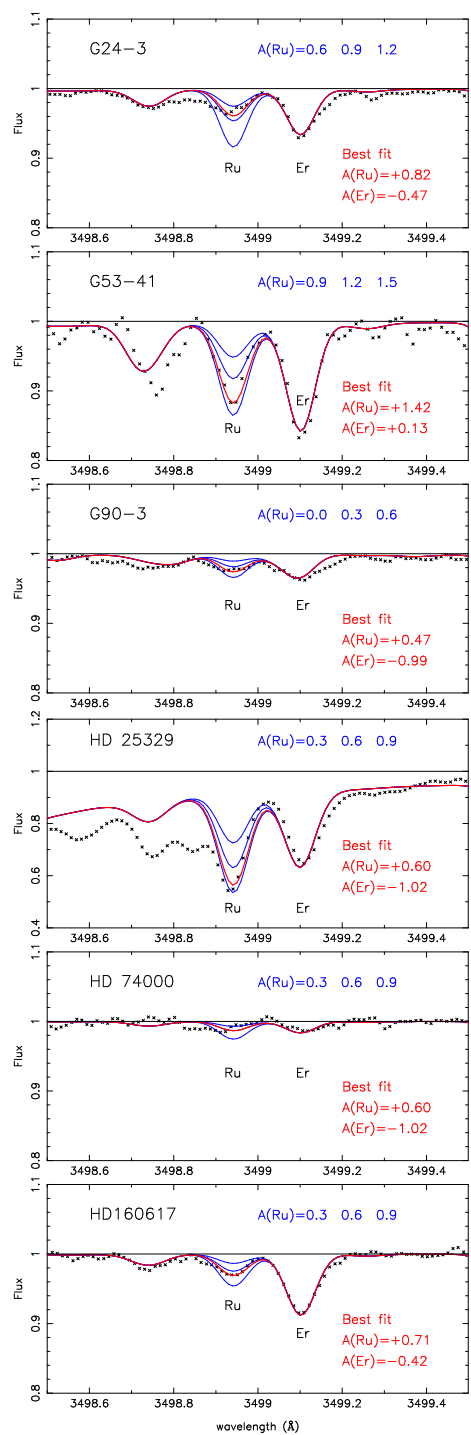

\begin{center}
\resizebox{6.0cm}{!}
{\includegraphics [clip=true]{G24-3-Ru.ps}}
\resizebox{6.0cm}{!}
{\includegraphics [clip=true]{G53-41-Ru.ps}}
\resizebox{6.0cm}{!}
{\includegraphics [clip=true]{G90-3-Ru.ps}}
\resizebox{6.0cm}{!}
{\includegraphics [clip=true]{HD25329-Ru.ps}}
\resizebox{6.0cm}{!}
{\includegraphics [clip=true]{HD74000-Ru.ps}}
\resizebox{6.0cm}{!}
{\includegraphics [clip=true]{HD160617-Ru.ps}}
\end{center}
\caption[]{Example of  fit of the observed and computed spectra in the region of the Ru and Er lines. The best fit is indicated by the red line.}
\label{Fig:spectra}
\end{figure}

\begin{figure*}
\begin{center}
\resizebox{6.0cm}{!}
{\includegraphics [clip=true]{abundSrFe.ps}}
\resizebox{6.0cm}{!}
{\includegraphics [clip=true]{abundYFe.ps}}
\resizebox{6.0cm}{!}
{\includegraphics [clip=true]{abundZrFe.ps}}
\resizebox{6.0cm}{!}
{\includegraphics [clip=true]{abundMoFe.ps}}
\resizebox{6.0cm}{!}
{\includegraphics [clip=true]{abundRuFe.ps}}
\resizebox{6.0cm}{!}
{\includegraphics [clip=true]{abundPdFe.ps}}
\resizebox{6.0cm}{!}
{\includegraphics [clip=true]{abundAgFe.ps}}
\resizebox{6.0cm}{!}
{\includegraphics [clip=true]{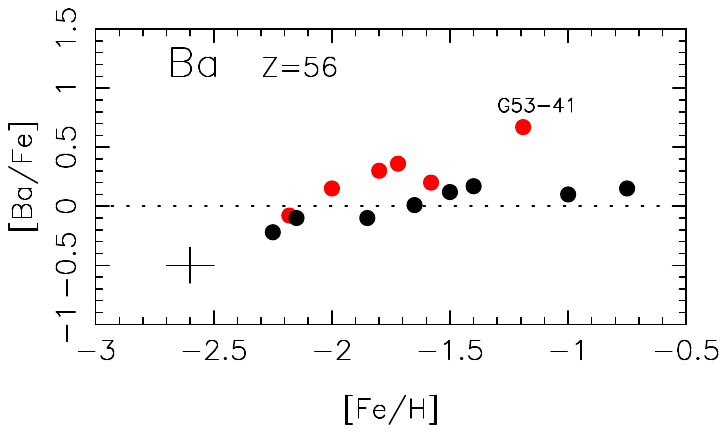}}
\resizebox{6.0cm}{!}
{\includegraphics [clip=true]{abundLaFe.ps}}
\resizebox{6.0cm}{!}
{\includegraphics [clip=true]{abundCeFe.ps}}
\resizebox{6.0cm}{!}
{\includegraphics [clip=true]{abundNdFe.ps}}
\resizebox{6.0cm}{!}
{\includegraphics [clip=true]{abundEuFe.ps}}
\resizebox{6.0cm}{!}
{\includegraphics [clip=true]{abundDyFe.ps}}
\resizebox{6.0cm}{!}
{\includegraphics [clip=true]{abundErFe.ps}}
\resizebox{6.0cm}{!}
{\includegraphics [clip=true]{abundYbFe.ps}}
\end{center}
\caption[]{[X/Fe]  vs. [Fe/H] for the N-rich dwarfs (red filled circles) and the normal dwarfs in the same interval of metallicity (black filled circles). The uncertainty on [Fe/H] and [X/Fe] is indicated by the black cross.}
\label{Fig:XFe}
\end{figure*}

\section{Observations and model atmosphere parameters}

The spectra of all the stars were mainly retrieved from the  ESO archives, (UVES, HARPS, FEROS, ESPRESSO spectra) or the  Keck HIRES archive. One OHP-Sophie spectrum was also used. All the spectra have a resolving power $ R \ge 40000 $ and a S/N  > 100 (see Paper I for  more detailed information about the spectra).

The stars studied in this work are mainly the same as   studied in Paper I.
However, in order to improve the comparison of the N-rich stars with `normal' metal-poor stars, we added two normal dwarfs in the interval -2<[Fe/H]<-1.5, HD\,108177 and HD\,218502.
The main parameters of the atmosphere of these stars were obtained in the same way as in Paper I.
The temperature and surface gravity of the stars were obtained by comparison of the Gaia DR3 photometry taking into account the distance of the star \citep{GaiaDR3-Vallenari22} and the reddening \citep{LallementBV18} with the theoretical values given by the PARSEC isochrones \citep{BressanMG12,MarigoGB17} computed at the metallicity of the star.\\
In this analysis we used MARCS model atmospheres \citep{GustafssonBE75,GustafssonEE03,Plez08} and the spectral synthesis code {\tt turbospectrum } \citep{AlvarezP98,Plez-code12}. The microturbulence velocity \vt\ was derived by requiring that the abundance deduced from individual \Feu~ lines be independent of the equivalent width of the line.\\
The atmospheric parameters of the stars and their metallicity [Fe/H] are given in Table \ref{Tab:AtmosParam}. Since the main goal of this work is the comparison of the abundances of the heavy elements in N-rich dwarfs and normal metal-poor dwarfs we  chose a sample of normal dwarfs with about the same atmospheric parameters as our sample of N-rich dwarfs. All these stars are turn-off stars with an effective temperature between 5900 and 6400K, and as a consequence we expect that the non-LTE effects are about the same in the N-rich stars and in the comparison stars. However, there is an exception for  the cool N-rich dwarf HD\,25329 with \Teff=4870K, but this very peculiar star is  discussed in Sect. \ref{sec25329}.

\begin{figure}
\begin{center}
\resizebox{6.0cm}{!}
{\includegraphics [clip=true]{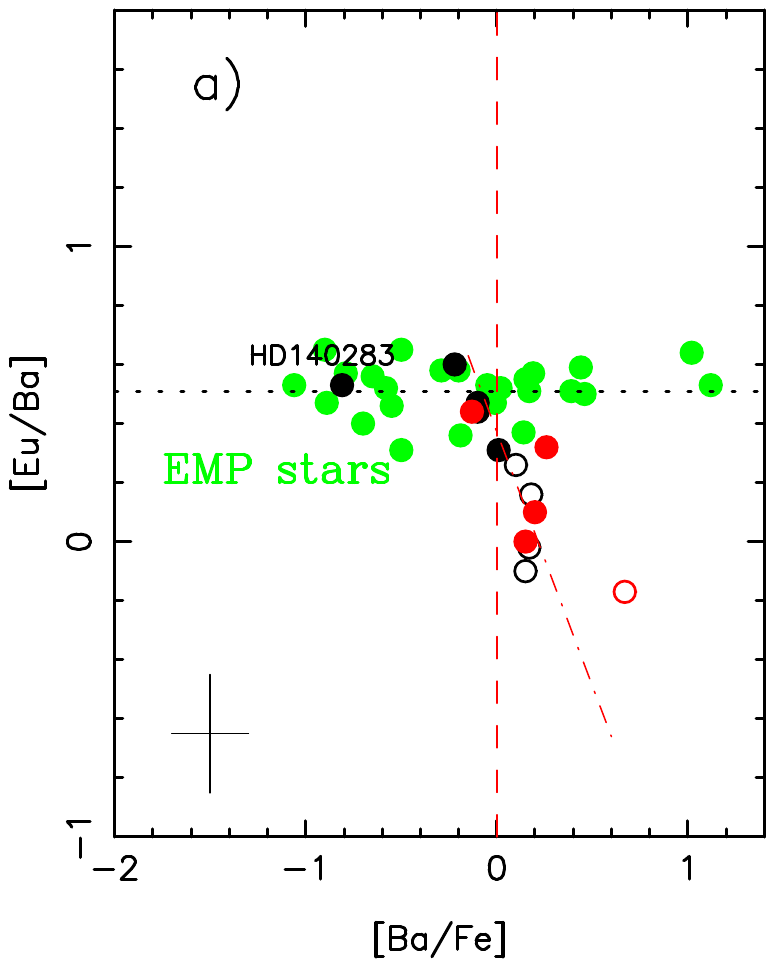}}
\resizebox{6.0cm}{!}
{\includegraphics [clip=true]{bafe-srLTE-Nrich.ps}}
\end{center}
\caption[]{Comparison of [Eu/Ba] and [Sr/Ba] vs. [Ba/Fe] in the intermediate metal-poor stars studied in this paper and in the EMP stars (green dots) from the literature. The black symbols represent the normal stars and the red symbols the N-rich stars; these symbols are open if   $\rm[Fe/H] \geq -1.5$ and  full if  $\rm -2.3 <[Fe/H]< -1.5$. The EMP stars (here with [Fe/H] < --2.5) are represented by green symbols. The plus signs (+)  represents the mean value of the abundance ratios in the subgiant and turn-off stars of the globular cluster 47\,Tuc ([Fe/H]=--0.7) and the crosses ($\times$)  in the giants of NGC\,6266  ([Fe/H]=--1.2). }
\label{Fig:SrBaFe}
\end{figure}

\begin{figure}
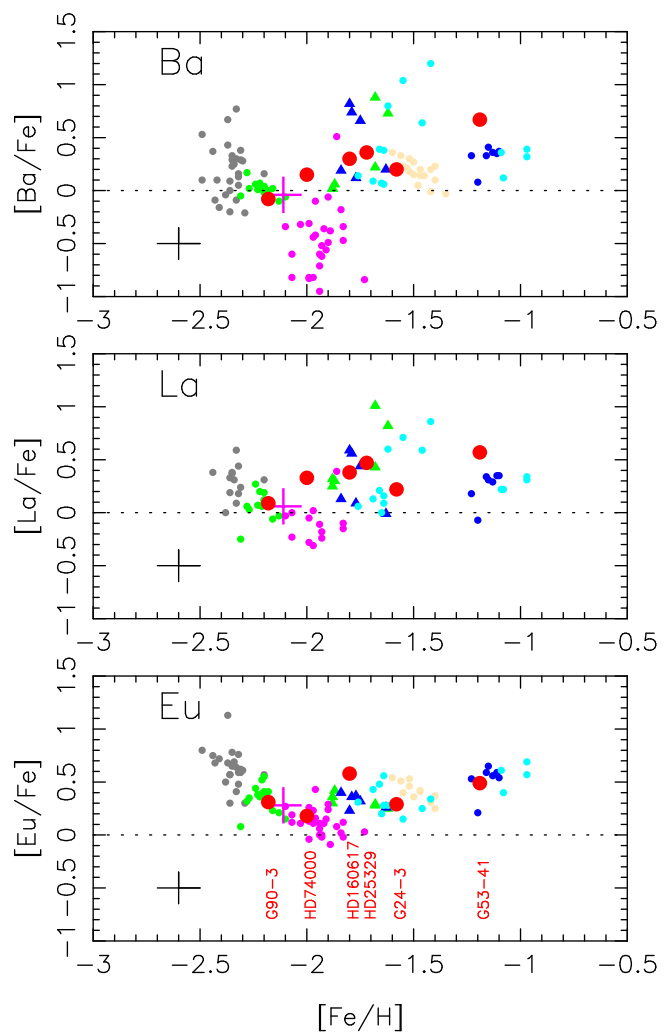

\begin{center}
\resizebox{8.5cm}{!}
{\includegraphics [clip=true]{abun-GC-BaFe.ps}}
\resizebox{8.5cm}{!}
{\includegraphics [clip=true]{abun-GC-LaFe.ps}}
\resizebox{8.5cm}{!}
{\includegraphics [clip=true]{abun-GC-EuFe.ps}}
\end{center}
\caption[]{Comparison of [Ba/Fe], [La/Fe], and [Eu/Fe] ratios in our  N-rich stars (red filled circles) with these ratios in M2 (blue triangles),  M15 (grey dots), NGC\,4833 (green dots), NGC\,5286 (green triangles), NGC\,5824 (pink dots), NGC\,6266 (blue dots), NGC\,6752 (light brown dots), and NGC\,7089 (cyan dots). The mean values of  [Ba/Fe], [La/Fe], and [Eu/Fe] measured by \citet{MucciarelliMB23} in NGC\,5824 stars are indicated by large pink crosses.}
\label{Fig:abClust}
\end{figure}

\section {Abundances of the heavy elements}   

We performed a standard local thermodynamic equilibrium (LTE) abundance analysis and, for all the heavy elements here studied, we derived the abundance line by line by comparing the  observed line profile with synthetic spectra computed with {\tt turbospectrum}. An example is given in Fig. \ref{Fig:spectra}.  Since the sample of stars here studied consists of metal-poor dwarfs or turn-off stars, most of the lines of the heavy elements are very weak in the linear part of the curve of growth ($\rm EW  \ll 50$ m\AA ), and are  thus not sensitive to hyperfine structure.  However, we took into account the hyperfine splitting of Eu, because in this case the shape of the line is strongly affected, and   of Ba, whose blue lines are rather strong. All the La and Nd lines are  weak,  and the hyperfine splitting can be neglected. \\
In the near UV we used rather strong \Yd~ and \Zrd~  lines (about 55 m\AA), but for these lines the atomic data of the levels are not available. 
As a test we computed the 420.5\,nm line of \Zrd\ with and without the hyperfine structure; the difference is less than 0.01 dex. Moreover, since no clear trend is observable between the blue and the near UV lines of \Zrd\ (see Table \ref{Tab:lines}), the hyperfine structure effect is probably not important for the near UV \Zrd~ lines in turn-off stars.\\
 In all the cases we adopted the  r-process isotopic fractions presented in \citet{SnedenCG08}. This distribution is not perfectly adapted when a star is partly enriched by the s-process \citep[see in particular ][]{RoedererMB16}, but we checked that, in fact, the total abundance is not very sensitive to the adopted fraction of the different isotopes. 

For each star the abundances of 15 n-capture elements from Sr to Yb are given in Table \ref{Tab:abund}.
The adopted solar abundances given in  this table  are from \citet{CaffauLS11} and \citet{LoddersPG09}. 
The  line-by-line abundances with the main characteristics of the lines are available in the  Appendix.

\section{Discussion}

In the early Galaxy the stars can only be enriched by different types of r-processes: weak r-process, strong solar type r-process, and an actinide boosted r-process \citep[see in particular][]{ArconesThielemann23}. 
Later, however,  at higher metallicity, the influence of the s-process contribution becomes important. 

In our Galaxy the main s-process contribution  indeed comes from LIM stars
in their AGB phase. They have a long evolutionary lifetimes, and thus contribute late in galactic evolution.
It generally appears at $\rm [Fe/H] \gtrsim -1.5 \pm 0.2$ dex  \citep[see e.g.][]{RoedererCK10}.
If a GC is formed from this material, the first-generation stars should   already be enriched in s-process elements by LIM AGB stars, for example  the field stars of this metallicity. The second-generation stars are formed from  intracluster material, and    later enriched  by massive AGB stars or fast-rotating massive stars  (FRMS; see e.g. \citealt{ArlandiniKW99} \citealt{PrantzosAC20}, \citealt{CowanSL21}).  In this case an s-process enrichment of the star does not necessarily mean that the matter that formed the star was enriched in s-process elements inside the cluster. 
On the contrary in a more metal-poor cluster ([Fe/H]<--1.5) an enrichment in s-process elements could indicate an enrichment within the cluster of the matter that formed the star.

Since Eu can only be synthesised  by the r-process, and Ba and La are predominantly s-process elements, even if they can also be produced by the r-process, the comparison of the abundances of Ba, La, and Eu is often used to determine the contribution of the s- and the r-process during the evolution of the Galaxy.

\begin{figure*}
\begin{center}
\resizebox{6.0cm}{!}
{\includegraphics [clip=true]{patHD140283.ps}}
\resizebox{6.0cm}{!}
{\includegraphics [clip=true]{patHD84937.ps}}
\resizebox{6.0cm}{!}
{\includegraphics [clip=true]{patG90-3.ps}}
\resizebox{6.0cm}{!}
{\includegraphics [clip=true]{patHD19445.ps}}
\resizebox{6.0cm}{!}
{\includegraphics [clip=true]{patHD74000.ps}}
\resizebox{6.0cm}{!}
{\includegraphics [clip=true]{patHD218502.ps}}
\resizebox{6.0cm}{!}
{\includegraphics [clip=true]{patHD160617.ps}}
\resizebox{6.0cm}{!}
{\includegraphics [clip=true]{patHD25329.ps}}
\resizebox{6.0cm}{!}
{\includegraphics [clip=true]{patHD108177.ps}}
\resizebox{6.0cm}{!}
{\includegraphics [clip=true]{patG24-3.ps}}
\resizebox{6.0cm}{!}
{\includegraphics [clip=true]{patHD166913.ps}}
\resizebox{6.0cm}{!}
{\includegraphics [clip=true]{patHD94028.ps}}
\resizebox{6.0cm}{!}
{\includegraphics [clip=true]{patG53-41.ps}}
\resizebox{6.0cm}{!}
{\includegraphics [clip=true]{patHD76932.ps}}
\resizebox{6.0cm}{!}
{\includegraphics [clip=true]{patHD97916.ps}}
\end{center}
\caption[]{Abundance patterns of the n-capture elements observed in our sample of stars sorted by [Fe/H]. The abundances of the elements are normalised to Eu.  The N-rich stars are indicated by red symbols and the normal stars by black symbols. Predictions of the main r-process following \citet{Wanajo07,BarbuySH11,SiqueiraSB13}, (hot and cold models) are indicated by blue lines. When these predictions did not match the observations between Ru and Yb and in particular if the Ba abundance is too high compared to the Eu abundance, the solar abundance pattern (green dashed line) has been added. The position of Ba and Eu are underlined by black arrows.}
\label{Fig:pattern}
\end{figure*}

\subsection{Comparison of [X/Fe] in   N-rich and in   normal stars}

In the first step we tried to compare the [X/Fe] ratios in the N-rich and normal dwarfs (Fig.\,\ref{Fig:XFe}), in the interval of metallicity. $\rm -2.3 <[Fe/H]< -0.7$, corresponding to the metallicity interval of our N-rich stars. A priori there is no strong difference between the behaviour of the N-rich stars and the normal stars.\\


In Fig.\,\ref{Fig:XFe} the scatter of the first peak elements from [Sr/Fe] to [Ru/Fe] is small, and we measured\\
<[Sr/Fe]>$\backsimeq$ 0.4, 
<[Y/Fe]>$\backsimeq$ 0.0, 
<[Zr/Fe]>$\backsimeq$ 0.35, 
<[Mo/Fe]>$\backsimeq$ 0.7,   
<[Ru/Fe]>$\backsimeq$ 0.7 and
<[Pd/Fe]>$\backsimeq$ 0.2.\\ 

The high abundance of Mo and Ru compared to the abundances of Sr, Y, Zr, and Pd, in intermediate metal-poor stars were pointed out by \citet{Peterson13}. This could indicate that these elements were mainly formed in the low-entropy domain of a high-entropy wind (HEW) in a type\,II  supernova \citep{FarouqiKP09,FarouqiKP10,Peterson11,Peterson13,FarouqiTR22}.

[Ba/Fe] and [La/Fe] seem statistically slightly higher in the N-rich stars than in normal stars, and this could indicate a more important contribution of the s-process in the N-rich stars. If the N-rich stars are, as expected, second-generation stars formed in globular clusters, these stars were formed from an intracluster gas polluted by H-process  material ejected by a first generation of massive AGB stars, which should be more Ba-rich than the gas that formed the first generation of stars.\\
We note that G53-41 (the least metal-poor star  in our sample) seems to be particularly rich in Ba and also La. This star, with $\rm[Fe/H] \simeq -1.2$, could have been formed in a GC built from  interstellar medium (ISM) already enriched in s-elements by LIM AGB stars, and later enriched inside the cluster by massive AGB or FRMS stars.

\subsection{Comparison of  [Eu/Fe], [Eu/Ba], and [Sr/Ba] in  extremely and intermediate metal-poor stars}

In Fig.\,\ref{Fig:SrBaFe} we compare the behaviour of the intermediate metal-poor stars studied in this work, to the extremely metal-poor  (EMP) stars with [Fe/H] < -2.5 \citep[e.g.][]{FrancoisDH07,SiqueiraHB14,SpiteSB18}
in two diagrams: [Eu/Ba] vs. [Ba/Fe] and [Sr/Ba] vs. [Ba/Fe].

In the EMP stars, at very low metallicity,  the scatter of the heavy elements is much larger than in intermediate metal-poor stars, [Eu/Fe] varies from about --0.3~ \citep[in e.g. HD\,122563 or HD\,88609;][]{HondaAI06,HondaAI07} to +1.6\,dex~ \citep[in e.g. CS\,22892-52;][]{SnedenMP96,SnedenCL03} (or CS\,31082-001) \citep{HillPC02,SiqueiraSB13,ErnandesCB03},
two well-known r-rich stars). The EMP stars are then classified as `r-poor' if [Eu/Fe]<--0.3, `normal' if $\rm-0.3 <[Eu/Fe] < +0.3$, and `r-rich' for higher values of this ratio (further classified as `r-I' if [Eu/Fe] is between +0.3 and +1, and `r-II' if [Eu/Fe]>+1;  \citealt{Beers-Chris05}).\\
In our sample of intermediate-metallicity stars  ($\rm -2.3 <[Fe/H]< -0.7$) the ratio [Eu/Fe] is about +0.3 dex in all the stars, in good agreement with the results of \citet{Peterson13}. The enrichment in r-process elements is stabilising (at the level of the r-I stars), since over time   a larger number of stars contributes to the heavy element  enrichment of the ISM  from which the stars are formed.

Since the matter that formed the EMP stars was enriched, at least for the second peak elements Ba and Eu, only through the r-process \citep [e.g.][]{PrantzosAC20,CowanSL21},  the ratio [Eu/Ba] (Fig. \ref{Fig:SrBaFe}a)  is practically constant and close to the theoretical r-process ratio value (black dotted line).  In the intermediate metal-poor stars, [Eu/Ba] decreases rapidly when [Ba/Fe] increases because the contribution  from  s-process elements ejected by LIM stars in their AGB phase becomes significant.\\ 
However it seems that  this s-process enrichment sometimes happens at lower metallicity in the N-rich stars. G24-3 and HD\,74000 with a metallicity of --1.6 and --2.0\,dex  have a low [Eu/Ba] ratio; in  Fig. \ref{Fig:SrBaFe}a) they are in a region where normally  only  stars with [Fe/H] > --1.5\,dex are found. \\
In HD\,140283, the r-poor EMP star in our reference sample, [Eu/Ba] is also typical of the r-process. The star has a very low value of [Ba/Fe] (never observed in intermediate-metallicity stars), but also a very low value of [Eu/Fe].

In the EMP stars there is no correlation between the abundance of Sr and the abundance of Ba  (Fig. \ref{Fig:SrBaFe}b); a star can be Ba-poor and Sr-rich. Ba is indeed  formed almost only by the main r-process, but Sr can   also be formed through a weak s-process in fast-rotating low-metallicity massive stars \citep{FrischknechtHT12,CescuttiCH13} or through a weak r-process \citep{FarouqiTR22}. In Fig. \ref{Fig:SrBaFe}b  almost all the EMP stars are located in a region delimited by a line representing the [Sr/Ba] ratio of the main r-process (black dotted line) and the line $\rm[Sr/Ba] \simeq 0.5 - [Ba/Fe]$ (green dashed line). On the contrary, in the intermediate metal-poor stars, the correlation between [Sr/Ba] and [Ba/Fe] is good, and all the dots (for normal and N-rich stars) are on a line $\rm[Sr/Ba] \simeq 0.4 - [Ba/Fe]$, which means that $\rm[Sr/Fe] \simeq 0.4$\,dex (see also Fig.\,\ref{Fig:XFe}). 

There is such a big difference in the behaviour of the heavy elements in the EMP stars and in the intermediate metal-poor stars that we have not included HD\,140283  in Fig.\,\ref{Fig:XFe} since our N-rich stars are not so metal poor. \\

\subsection{Comparison of abundance ratios in the N-rich stars and in the globular cluster stars}

Following \citet{RoedererCK10}, in the metal-poor globular clusters with [Fe/H]<--1.4 the abundance ratios of the neutron-capture elements suggest that significant amounts of s-process material were not present in the ISM from which these cluster stars were formed. In Fig.\,\ref{Fig:SrBaFe}a and b  we   include two more metal-rich globular clusters: NGC\,6266 with $\rm[Fe/H] \simeq -1.2$ and 47 Tuc $\rm[Fe/H] \simeq -0.7$.
The position of these intermediate metal-poor clusters is, as expected, close to the position of the intermediate metal-poor field stars. Globular cluster stars and field stars of intermediate metallicity were thus formed from ISM of similar compositions.

In Fig. \ref{Fig:abClust} we compare the behaviour of [Ba/Fe], [La/Fe], and [Eu/Fe] in our sample of dwarf N-rich stars with these ratios in seven globular clusters where at least the abundance of Ba and Eu were measured for several stars: M15 \citep{SobeckKS11,WorleyHS13}, NGC\,4833 \citep{RoedererThompson15}, NGC\,5286 \citep{MarinoMK15}, NGC\,5824 \citep{RoedererMB16}, NGC\,6752 \citep{JamesFB04a,JamesFB04b}  NGC\,7089 \citep{YongRG14}, and NGC\,6266 \citep{YongABDC14}.
 Unfortunately, the nitrogen abundance of the globular cluster stars in Fig. \ref{Fig:abClust} is generally unknown, and 
when it exists it is based on the CN violet band. The CN violet band  is well known to be affected by NLTE and 3D effects, which  have not been satisfactorily modelled to date.  A homogeneous study of the N abundance in globular cluster stars based on the NH UV band is clearly badly needed.
The observed ratios in our sample of N-rich stars is compatible with the ratios observed in globular cluster stars. 
HD\,25329 seems to have a rather high [Eu/Fe] ratio, but such a high value of [Eu/Fe] is also observed in M15 stars.\\  
We note that, following \citet{RoedererMB16}, NGC\,5824 has a very low value of [Ba/Fe]: <[Ba/Fe]> $\approx -0.6$\,dex. In the other GCs  and in the field stars (Fig.\,\ref{Fig:abClust} and Fig.\,\ref{Fig:XFe}), the mean value of [Ba/Fe] is close to zero or higher. \citet{RoedererMB16} remark that the value of the Ba abundance they measured in this cluster is uncertain because it is derived from the single 455.4\,nm line, which is highly sensitive to the choice of the microturbulence velocity. 
In a recent paper on    globular clusters in the Small Magellanic Cloud, \citet{MucciarelliMB23} give the average abundance of Ba, La, and Eu in  six stars of  NGC\,5824. These measurements were done from UVES spectra  centred in the visible region, and the mean value is indicated in Fig.\,\ref{Fig:abClust} by a  pink cross. For La and Eu their abundance values are in agreement with the values found by \citet{RoedererMB16}, but the mean value of [Ba/Fe] is significantly higher than the values measured by \citet{RoedererMB16}.
They found [Ba/Fe] = --0.04 with a very small dispersion, $\sigma=0.05$, in good agreement with the values found in the other GCs and in the field stars.

\subsection {Pattern of heavy element abundances}
The pattern of element abundances is presented in  Fig. \ref{Fig:pattern} for all the stars in our sample. It provides an overview of the abundances of the n-capture elements in each star. These abundances are normalised to Eu. Since we expect that in normal stars the pattern is similar to the r-process pattern at low metallicity,  and that it  then evolves towards an s-process enriched pattern as soon as $\rm[Fe/H]\simeq-1.5$,  the stars are ordered by increasing metallicity. The abundances of normal stars are represented by black symbols and of N-rich stars by red symbols. The blue lines represent the predictions of Wanajo for the hot and cold models of pure r-process \citep{Wanajo07,BarbuySH11,SiqueiraSB13}, and the green line the solar abundance pattern. We expect that, in normal stars, at low metallicity the pattern of element abundance is close to the r-process predictions, and that at higher metallicity this pattern tends to the solar abundance pattern.\\
The abundance patterns of the N-rich stars could show the influence of an enrichment by more massive AGB inside the globular cluster. A good example is HD\,74000, which has an abundance pattern very close to the solar one, despite its low metallicity.\\ 
Moreover, if some N-rich stars were formed from  gas already enriched  by the first low-mass AGB,  and further, inside a cluster, by the ejecta of the first more massive AGB, we could expect that some N-rich stars are particularly Ba-rich. However none of our N-rich stars has an abundance pattern above the solar one.\\ 
In contrast, the abundance patterns of three N-rich stars, HD\,25329, HD\,160617, and G\,90-3, are compatible with a formation of the n-capture elements by a pure r-process. As a consequence, a star can be strongly enriched in nitrogen and not significantly in s-process elements.\\ 
On the other hand, we note that in all the stars (normal or N-rich field stars) the elements heavier than Eu are always compatible with a formation by the main r-process.\\
A rather strong Pb line has been observed in one star of NGC\,5824 by \citet{RoedererMB16}.
The presence of Pb is the signature of the s-process in the  low-metallicity AGB stars. Due to the scarcity of seed nuclei, the s-process indeed shifts to produce the heaviest elements, such as  Pb \citep{BisterzoGS10}. In none of our N-rich stars   could we detect the Pb line.

\begin{figure}
\begin{center}
\resizebox{7.0cm}{!}
{\includegraphics [clip=true]{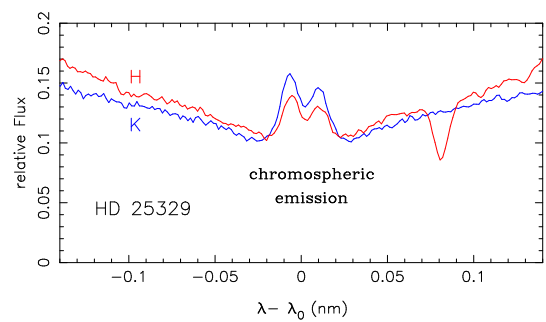}}
\end{center}
\caption[]{Profile of the botttom of the H and K lines of Ca\,II in HD\,25329. The red line represents the H line and the blue line the K line.}
\label{Fig:chrom}
\end{figure}

\section {HD\,25329, a very peculiar metal-poor N-rich dwarf} \label{sec25329}

HD\,25329 is the coolest N-rich metal-poor dwarf in our sample (about 4900K). Its metallicity is [Fe/H]=--1.72. 
\citet{BarbuySS87}~  measured in its atmosphere the magnesium isotopic composition. They found that it has a high proportion of $\rm ^{24}Mg$, as do the other metal-poor stars with this metallicity, higher than it is in the more metal-rich and thus younger stars. Therefore, HD\,25329 was formed from a material that has characteristics of the old populations of our Galaxy, and thus we expected it to be about 12 Gyr old, like most of the globular clusters with a metallicity close to [Fe/H]=--1.5.\\  
In Paper\,I we tried to determine the age of this star by a direct comparison of the Parsec isochrones to Gaia photometry in a diagram G versus $\rm G _{BP}- G_{RP}$, but we could not because the star is still on the main sequence. However \citet{CasagrandeSA11} from a Bayesian analysis based on the Geneva-Copenhagen survey \citep{NordstromMA04,HolmbergNA09} and Str\"omgren photometry \citep{Stromgren87}, was able to  estimate that this star is  about 5\,Gyr  old. As a consequence, it seems to be very young for its metallicity.\\ 
HD\,25329 does not present any signs of binarity: the Gaia error on the radial velocity is only 0.12\,\kms\ over 16 transits and the photometric error on the G magnitude is only $0.18~10^{-3}$ magnitude over 440 transits.

On the other hand, HD\,25329 is known to have a strong chromospheric activity (see the profiles of the H and K Ca\,II lines in Fig. \ref{Fig:chrom}). Following \citet{Pace13},  strong activity is a characteristic of a young galactic population.\\ 
It has been shown that the metal-poor dwarfs, even if they are very old, have an active chromosphere \citep{PetersonS97,SpitePG17}. However, this activity is weak, it is never detected in the \Cad\ lines,   and only becomes   visible in the ultraviolet \Mgd\ lines. The background continuum of a solar-type star is indeed lower in the ultraviolet than in the optical. Moreover, magnesium is much more abundant than calcium, and thus even a weak chromospheric activity becomes detectable in the UV  \Mgd\ lines.

The question now arises of how we can  reconcile the strong chromospheric activity of HD\,25329 and its very low metallicity. 
One possibility is that this star is truly
young, but that it comes from a globular cluster of a dwarf galaxy with a low star formation rate, which was later accreted by our Galaxy \citep[see e.g.][]{SkuladottirHC20}. However, it should be kept in mind that there are no known metal-poor globular clusters that are this young.

 Another possibility is that this star is an old star, as expected from its metallicity, but that it is, like the classical blue-stragglers, the result of the merging of two stars, but two very low-mass stars. \citet{RyanBK01} suggest that such stars would not yet appear bluer than the turn-off, but once the main sequence turn-off reaches lower mass these objects would lag the evolution of normal stars and appear bluer.
 Following \citet{RyanBK01} HD\,25329 could be thus a `blue-straggler-to-be'. 
In a classical blue-straggler  (like HD\,97916; see Paper I), Li is ultra-deficient, the line is not visible, but Be is also strongly deficient. Be is also a fragile element, but it is destroyed at a higher temperature than Li, at about $\rm 3.5\, 10^6\,K$, and thus its abundance is not as temperature-sensitive as that of Li. It seems that the Be abundance in HD\,25329 is normal for its metallicity, but  since the Be lines are embedded in the strong OH band, this result is not completely reliable, and as a consequence we cannot completely rule out this hypothesis.

\section {Conclusion}
We studied the behaviour of the neutron capture elements in a sample of six field N-rich stars suspected to be formed as second-generation stars in GCs (see Paper I). Our conclusions can be summarized as follows: 

\noindent-- When the ratios [X/Fe] of the first peak elements from Sr to Ag are compared in the normal stars and in the N-rich stars, no significant difference is found. The scatter of the abundance ratios is very small  for the first peak elements, and in particular for Sr, Y, Zr, Mo, and Ru.  The scatter of the ratios  [X/Fe]  of the second peak elements, from Ba to Yb, is more important, and it seems that the [Ba/Fe] ratio is statistically higher in the field N-rich stars suggesting an s-process enrichment of these stars.

\noindent-- Generally speaking, there is a very strong difference between the behaviour of the EMP stars and the intermediate metal-poor stars. In the EMP stars the ratio [Eu/Ba] is almost constant and independent of the r process enrichment. Both elements are indeed formed by the main-r process. In contrast, in moderately metal-poor stars, [Eu/Ba] decreases linearly as [Ba/Fe] increases. This decrease begins at around [Fe/H]=--1.5 in normal stars, marking the arrival of pollution by s-process material ejected by low-mass AGB stars. 

\noindent-- In the N-rich stars G24-3 and HD\,74000, this pollution by s-process material is already visible at  [Fe/H]=--1.6 and --2.0 suggesting that these stars have been over-enriched in s-process material by, for example, massive AGB stars (or FRMS stars) inside a globular cluster.
A possible alternative interpretation is that
these N-rich stars have escaped from globular clusters that were formed in dwarf
galaxies, where the s-process enrichment starts at lower metallicities  
\citep[see e.g.][]{SkuladottirHC20}. 
However, this interpretation is not fully consistent
with the ages of our stars. \citet{BK2023b} have argued that clusters formed in situ
and that they  accreted, and thus formed in dwarf galaxies, are displaying
two distinct age--[Fe/H] sequences. The accreted clusters are about 1\,Gyr
younger than  in situ clusters at any given metallicity.
With the exception of HD\,25329 our stars are older than 12\,Gyr, 
which would place them in the  in situ age--[Fe/H] sequence.

\noindent--  In moderately metal-poor stars the large scatter of [Sr/Ba] observed in EMP stars disappears,
 $\rm[Sr/Ba] \simeq 0.4 - [Ba/Fe],$ which means  that [Sr/Fe] is almost constant and is equal to 0.4\,dex with a very small scatter. 

\noindent--  The abundance ratios of n-capture elements in intermediate metal-poor field stars and globular cluster stars are very similar. This suggests that they were formed from  ISM with   similar compositions.  
In about half of our sample of N-rich stars the abundance pattern of the n-capture elements is compatible with an enrichment in s-process elements that is unexpected at this metallicity (\mbox{G24-3}, HD74000, and maybe HD160617). G53-41 is also s-process enriched, but, due to its relatively high metallicity, we cannot decide wether this enrichment happened before the formation of the cluster or inside the cluster.
Two very metal-poor N-rich stars, G90-3 and HD25329, have an abundance pattern compatible with a pure r-process formation.

\noindent-- In the frame of this work we have brought to light a very peculiar N-rich star: HD\,25329. This star is known for its high chromospheric activity. It seems to be rather young, but it is very metal-poor ([Fe/H]=--1.7).
We suggest that it could have been accreted from a globular cluster formed  in a dwarf galaxy with a low star formation rate.

\begin {acknowledgements} 
This work uses results from the European Space Agency (ESA) space 
mission Gaia.  Gaia data are being processed by the Gaia Data
Processing and Analysis Consortium (DPAC).  Funding for the DPAC is
provided by national institutions, in particular the institutions
participating in the Gaia MultiLateral Agreement (MLA).  The Gaia
mission website is https://www.cosmos.esa.int/gaia.  The Gaia archive
website is https://archives.esac.esa.int/gaia.  
\end{acknowledgements}

{}

\appendix
\section{}

\begin{table*}
\begin{center}   
\caption[]{ 
 Characteristics of the lines and element abundances. In Cols. 4 to 9 are given the abundances for the N-rich stars and then in Cols. 10 to 14 the abundances of the new comparison stars.}
\label{Tab:lines}
\centering
\begin{tabular}{r@{~~}r@{~~}r@{~~~~~}c@{~~}c@{~~}c@{~~}c@{~~}c@{~~}c@{~~~~~~}c@{~~}c@{~~}c@{~~}c@{~~}c@{~~}rrr}
\hline
       &           &        &    HD25329&    HD74000&  HD160617&      G24-3&     G53-41&      G90-3&     HD76932&    HD97916& HD108177&  HD166913&   HD218502   \\
       &           &        &     N-rich&     N-rich&    N-rich&     N-rich&     N-rich&     N-rich&                                                            \\   
       &           &        &           &           &          &           &           &           &            &           &          &           &            \\
   Elem& lambda (\AA)&\loggf&       A(X)&       A(X)&      A(X)&       A(X)&       A(X)&       A(X)&        A(X)&       A(X)&      A(X)&       A(X)&      A(X)  \\
  Sr II&   4077.710&   0.167&       1.55&       1.39&      1.39&       1.48&       1.95&       0.96&        2.20&       2.24&      1.55&       1.76&       1.39 \\
  Sr II&   4215.520&  -0.145&       1.50&       1.42&      1.29&       1.53&       1.93&       0.98&        2.20&       2.23&      1.66&       1.83&       1.45 \\
       &           &        &           &           &          &           &           &           &            &           &          &           &            \\
   Y II&   3600.732&   0.280&          -&       0.33&      0.32&       0.70&       1.18&      -0.09&        1.27&       1.45&      0.61&       0.83&       0.38 \\
   Y II&   3601.916&  -0.180&          -&       0.33&      0.32&       0.63&       1.24&      -0.07&        1.28&         - &      0.58&       0.83&       0.38 \\
   Y II&   3611.043&   0.110&          -&       0.22&      0.20&       0.58&       1.19&      -0.21&        1.25&         - &      0.51&       0.71&       0.25 \\
   Y II&   3710.287&   0.523&          -&       0.21&      0.25&       0.63&       1.14&         - &        1.18&       1.34&      0.52&       0.76&       0.28 \\
   Y II&   3774.330&   0.291&          -&       0.27&        - &       0.65&       1.19&      -0.16&        1.17&       1.27&      0.55&       0.77&       0.32 \\
   Y II&   3788.694&   0.012&          -&       0.30&      0.36&       0.64&       1.21&      -0.11&        1.25&         - &      0.59&       0.76&       0.32 \\
   Y II&   3818.341&  -0.980&          -&       0.38&      0.32&       0.58&       1.13&      -0.20&        1.28&       1.33&      0.61&       0.84&         -  \\
   Y II&   3950.349&  -0.490&       0.61&       0.28&      0.40&         - &       1.18&      -0.12&        1.39&         - &        - &       0.86&         -  \\
   Y II&   4398.013&  -1.000&       0.67&       0.22&      0.33&       0.52&       1.10&      -0.24&        1.24&       1.34&      0.51&       0.78&       0.23 \\
   Y II&   4682.324&  -1.510&       0.77&       0.20&      0.30&       0.62&         - &      -0.24&        1.47&         - &      0.76&       0.84&         -  \\
   Y II&   4883.684&   0.070&       0.83&       0.30&      0.41&       0.64&       1.18&         - &        1.33&       1.49&      0.60&       0.85&       0.37 \\
   Y II&   5087.419&  -0.170&       0.61&       0.24&      0.35&       0.54&       1.12&      -0.18&        1.31&       1.39&      0.48&       0.79&       0.22 \\
   Y II&   5205.722&  -0.340&           &       0.29&      0.40&       0.57&       1.13&      -0.30&        1.35&       1.37&      0.51&       0.81&       0.30 \\
       &           &        &           &           &          &           &           &           &            &           &          &           &            \\
  Zr II&   3438.231&   0.410&          -&       0.99&      1.02&          -&       1.85&         - &          - &       1.46&      1.28&       1.26&       1.00 \\
  Zr II&   3496.205&   0.260&          -&       1.03&      1.04&       1.27&       1.86&       0.69&        1.76&       1.82&      1.36&       1.44&       1.06 \\
  Zr II&   3551.951&  -0.360&          -&       1.10&      1.07&       1.33&       1.98&       0.73&        2.00&       1.97&      1.38&       1.52&       1.11 \\
  Zr II&   3556.594&   0.070&          -&       1.00&      0.96&       1.21&       1.84&       0.60&          - &         - &      1.25&       1.38&       1.01 \\
  Zr II&   3576.853&  -0.120&          -&       1.04&      1.13&       1.32&       1.90&       0.69&        1.84&       1.90&        - &       1.49&       1.08 \\
  Zr II&   3614.765&  -0.252&          -&       1.01&      1.05&       1.17&       2.10&       0.61&        2.27&         - &      1.26&       1.46&       1.04 \\
  Zr II&   3836.761&  -0.120&          -&       0.96&      1.00&       1.27&       1.71&       0.61&        1.86&       1.87&      1.36&       1.44&       1.04 \\
  Zr II&   4149.217&  -0.030&       1.54&       1.10&      1.12&       1.35&       1.96&       0.63&        1.99&         - &      1.37&       1.59&       1.12 \\
  Zr II&   4161.213&  -0.720&       1.42&       1.15&      1.13&       1.35&       1.90&       0.65&        2.03&       2.00&      1.36&       1.59&       1.10 \\
  Zr II&   4208.977&  -0.460&          -&       1.10&      1.13&       1.36&       1.90&       0.65&        2.02&       2.04&      1.34&       1.58&       1.13 \\
       &           &        &           &           &          &           &           &           &            &           &          &           &            \\
   Mo I&   3864.103&  -0.010&     1.14: &       0.62&      0.75&       0.94&       1.42&       0.28&        1.56&        -  &      1.09&       1.16&       0.86 \\
       &           &        &           &           &          &           &           &           &            &           &          &           &            \\
   Ru I&   3498.394&   0.310&       0.83&       0.60&      0.71&       0.82&       1.42&       0.47&        1.43&         - &      0.80&       1.06&       0.61 \\
       &           &        &           &           &          &           &           &           &            &           &          &           &            \\
   Pd I&   3404.579&   0.320&       0.37&  $\le$0.0 &      0.04&       0.01&       0.81&      -0.51&        0.86&         - &      0.22&       0.54&      -0.03 \\
       &           &        &           &           &          &           &           &           &            &           &          &           &            \\
   Ag I&   3382.000&  -0.377& $\le$-0.62&     -0.20:&     -0.40& $\le$-0.4 &       0.32&$\le$-0.94 &        0.43&         - &     -0.11&      -0.10&     -0.44: \\
       &           &        &           &           &          &           &           &           &            &           &          &           &            \\
  Ba II&   4554.000&   0.170&       0.88&       0.25&      0.71&       0.71&         - &      -0.09&        1.36&       1.55&      0.59&       0.75&         -  \\
  Ba II&   4934.000&  -0.150&       0.71&       0.22&      0.64&         - &       1.50&         - &        1.25&       1.45&      0.47&       0.66&         -  \\
  Ba II&   5853.670&  -1.010&          -&       0.29&      0.57&       0.75&       1.55&         - &        1.42&       1.42&      0.53&       0.81&       0.22 \\
  Ba II&   6141.710&  -0.070&       0.76&       0.38&      0.70&       0.83&       1.71&         - &        1.50&       1.66&      0.52&       0.85&       0.14 \\
  Ba II&   6496.000&  -0.407&       0.88&       0.46&      0.72&       0.88&       1.82&         - &        1.55&       1.79&      0.53&       0.90&       0.29 \\
       &           &        &           &           &          &           &           &           &            &           &          &           &            \\
  La II&   3988.520&   0.170&      -0.21&      -0.55&     -0.36&      -0.25&       0.48&      -0.98&        0.38&       0.55&     -0.37&      -0.27&      -0.56 \\
  La II&   4086.709&  -0.032&      -0.07&      -0.55&     -0.30&      -0.21&       0.49&      -0.98&        0.43&       0.50&     -0.37&      -0.22&      -0.51 \\
  La II&   4123.218&   0.067&      -0.04&      -0.50&     -0.18&      -0.19&       0.59&      -0.89&        0.58&       0.32&     -0.20&      -0.13&      -0.65 \\
       &           &        &           &           &          &           &           &           &            &           &          &           &            \\
  Ce II&   4073.474&   0.210&       0.67&      -0.13&      0.16&       0.23&       0.81&      -0.53&        0.70&       0.93&      0.13&       0.15&      -0.30 \\
  Ce II&   4083.222&   0.270&       0.35&      -0.15&      0.28&       0.22&       0.97&      -0.29&        0.85&       1.25&      0.46&       0.29&         -  \\
  Ce II&   4137.645&   0.400&          -&      -0.04&      0.16&       0.02&       0.84&         - &        0.85&       0.81&      0.15&       0.30&      -0.07 \\
  Ce II&   4222.597&  -0.150&       0.74&      -0.05&      0.19&       0.29&       0.86&      -0.60&        0.81&       1.01&      0.06&       0.31&      -0.17 \\
  Ce II&   4486.909&  -0.180&       0.40&      -0.20&      0.19&       0.14&       0.86&         - &        0.77&       0.40&      0.24&       0.16&         -  \\
  Ce II&   4562.359&   0.210&         - &      -0.02&      0.10&       0.27&         - &      -0.58&        0.79&       0.87&      0.11&       0.20&         -  \\
       &           &        &           &           &          &           &           &           &            &           &          &           &            \\
  Nd II&   3826.410&  -0.410&         - &       0.08&     -0.13&        -  &       0.68&      -0.74&        1.08&         - &      0.19&      -0.03&        -   \\
  Nd II&   3863.410&  -0.040&         - &       0.03&        - &        -  &        -  &      -0.30&        0.80&         - &     -0.10&       0.09&         -  \\
  Nd II&   4021.330&  -0.100&       0.26&         - &        - &       0.28&       0.63&         - &          - &         - &        - &       0.13&         -  \\
  Nd II&   4061.080&   0.550&       0.44&         - &        - &       0.21&         - &         - &          - &         - &        - &         - &       0.16 \\
  Nd II&   4109.448&   0.350&          -&      -0.21&      0.02&       0.19&         - &      -0.65&        0.85&       0.80&     -0.05&       0.21&       0.20 \\
  Nd II&   4446.380&  -0.350&       0.21&      -0.17&      0.02&       0.19&       0.63&         - &        0.73&       1.00&      0.30&       0.03&         -  \\
       &           &        &           &           &          &           &           &           &            &           &          &           &            \\
  Eu II&   3724.930&  -0.090&         - &         - &     -0.70&          -&         - &         - &          - &        -  &        - &         - &         -  \\
  Eu II&   3819.670&   0.485&         - &      -1.28&     -0.70&      -0.79&      -0.21&      -1.38&         -  &       -   &        - &         - &         -  \\
  Eu II&   4129.720&   0.205&     -0.35:&      -1.32&     -0.71&      -0.74&      -0.14&      -1.32&        0.00&      -0.18&     -0.81&      -0.64&      -0.99 \\
       &           &        &           &           &          &           &           &           &            &           &          &           &            \\
  Dy II&   3407.800&   0.180&          -&          -&         -&       0.09&       0.70&      -0.53&        0.67&       0.65&     -0.08&      -0.03&      -0.33 \\
  Dy II&   3531.710&   0.770&      -0.29&      -0.94&     -0.11&      -0.21&       0.59&      -0.92&        0.66&       0.57&     -0.20&      -0.16&      -0.45 \\
  Dy II&   3536.020&   0.530&      -0.21&         - &     -0.12&      -0.20&       0.52&      -0.80&        0.58&       0.50&     -0.24&      -0.11&      -0.33 \\
  Dy II&   3563.000&  -0.360&          -&         - &         -&      -0.17&       0.46&      -0.54&        0.59&       0.62&     -0.33&       0.02&         -  \\
  Dy II&   3694.308&  -0.660&          -&         - &     -0.11&      -0.17&       0.48&      -0.79&        0.63&         - &        - &       0.01&         -  \\
  Dy II&   3944.680&   0.110&          -&      -0.92&         -&         - &         - &         - &          - &         - &        - &      -0.07&         -  \\
       &           &        &          -&           &          &           &           &           &            &           &          &           &            \\
  Er II&   3499.103&   0.290&       0.00&      -1.02&     -0.42&      -0.47&       0.13&      -0.99&        0.45&       0.14&     -0.51&      -0.25&      -0.77 \\
  Er II&   3692.649&   0.280&      -0.12&      -0.77&     -0.30&      -0.39&       0.30&      -0.99&        0.52&       0.35&     -0.29&      -0.27&      -0.55 \\
       &           &        &           &           &          &           &           &           &            &           &          &           &            \\
  Yb II&   3694.190&  -0.320&      -0.34&      -0.87&     -0.34&      -0.37&       0.60&      -1.22&        0.77&       0.57&     -0.46&      -0.27&      -0.70 \\
\hline    
\end{tabular}  
\end{center}   
\end{table*}


\begin{thebibliography}{}

\bibitem[Alvarez \& Plez(1998)]{AlvarezP98}
Alvarez R., Plez B., 1998, A\&A 330, 1109

\bibitem[Arcones \& Thielemann(2023)]{ArconesThielemann23} 
Arcones, A. \& Thielemann, F.-K.\ 2023, \aapr, 31, 1. doi:10.1007/s00159-022-00146-x

 
\bibitem[Arlandini et al.(1999)]{ArlandiniKW99} 
Arlandini, C., K{\"a}ppeler, F., Wisshak, K., et al.\ 1999, \apj, 525, 886. doi:10.1086/307938

\bibitem[Arnould \& Goriely(2020)]{ArnouldGoriely20} 
Arnould, M. \& Goriely, S.\ 2020, Progress in Particle and Nuclear Physics, 112, 103766. doi:10.1016/j.ppnp.2020.103766

\bibitem[Barbuy et al.(1987)]{BarbuySS87} 
Barbuy, B., Spite, F., \& Spite, M.\ 1987, \aap, 178, 199

\bibitem[Barbuy et al.(2011)]{BarbuySH11} 
Barbuy, B., Spite, M., Hill, V., et al.\ 2011, \aap, 534, A60. doi:10.1051/0004-6361/201117450

\bibitem[Beers \& Christlieb(2005)]{Beers-Chris05} 
Beers, T.~C. \& Christlieb, N.\ 2005, \araa, 43, 531. doi:10.1146/annurev.astro.42.053102.134057


\bibitem[Belokurov \& Kravtsov(2023)]{BK2023b} Belokurov, V. \& Kravtsov, A.\ 2023, arXiv:2309.15902. doi:10.48550/arXiv.2309.15902

\bibitem[Bisterzo et al.(2010)]{BisterzoGS10} 
Bisterzo, S., Gallino, R., Straniero, O., et al.\ 2010, \mnras, 404, 1529. doi:10.1111/j.1365-2966.2010.16369.x

\bibitem[Bressan et al.(2012)]{BressanMG12} 
Bressan, A., Marigo, P., Girardi, L., et al.\ 2012, \mnras, 427, 127 

\bibitem[Caffau et al.(2011)]{CaffauLS11} 
Caffau, E., Ludwig, H.-G., Steffen, M., et al.\ 2011, \solphys, 268, 255. doi:10.1007/s11207-010-9541-4

\bibitem[Casagrande et al.(2011)]{CasagrandeSA11} 
Casagrande, L., Sch{\"o}nrich, R., Asplund, M., et al.\ 2011, \aap, 530, A138. doi:10.1051/0004-6361/201016276

\bibitem[Cescutti et al.(2013)]{CescuttiCH13} 
Cescutti, G., Chiappini, C., Hirschi, R., et al.\ 2013, \aap, 553, A51. doi:10.1051/0004-6361/201220809

\bibitem[Cowan et al.(2021)]{CowanSL21} 
Cowan, J.~J., Sneden, C., Lawler, J.~E., et al.\ 2021, Reviews of Modern Physics, 93, 015002. 
doi:10.1103/RevModPhys.93.015002

\bibitem[Ernandes et al.(2023)]{ErnandesCB03}
Ernandes, H., Castro, M., Barbuy, B., et al.\ 2023, \mnras, doi:10.1093/mnras/stad1764

\bibitem[Farouqi et al.(2009)]{FarouqiKP09} 
Farouqi, K., Kratz, K.-L., \& Pfeiffer, B.\ 2009, \pasa, 26, 194. doi:10.1071/AS08075

\bibitem[Farouqi et al.(2010)]{FarouqiKP10} 
Farouqi, K., Kratz, K.-L., Pfeiffer, B., et al.\ 2010, \apj, 712, 1359. doi:10.1088/0004-637X/712/2/1359

\bibitem[Farouqi et al.(2022)]{FarouqiTR22} 
Farouqi, K., Thielemann, F.-K., Rosswog, S., et al.\ 2022, \aap, 663, A70. doi:10.1051/0004-6361/202141038

\bibitem[Fran{\c{c}}ois et al.(2007)]{FrancoisDH07} 
Fran{\c{c}}ois, P., Depagne, E., Hill, V., et al.\ 2007, \aap, 476, 935

\bibitem[Frischknecht et al.(2012)]{FrischknechtHT12} 
Frischknecht, U., Hirschi, R., \& Thielemann, F.-K.\ 2012, \aap, 538, L2. doi:10.1051/0004-6361/201117794

\bibitem[Frischknecht et al.(2016)]{FrischknechtHP16} 
Frischknecht, U., Hirschi, R., Pignatari, M., et al.\ 2016, \mnras, 456, 1803. doi:10.1093/mnras/stv2723

\bibitem[Gaia Collaboration et al.(2022)]{GaiaDR3-Vallenari22} 
Gaia Collaboration, Vallenari, A., Brown, A.~G.~A., et al.\ 2022, arXiv e-prints, aXiv:2208.00211

\bibitem[Gustafsson et al.(1975)]{GustafssonBE75}
Gustafsson B., Bell R. A., Eriksson K., Nordlund \AA., 1975, A\&A, 42, 407 

\bibitem[Gustafsson et al.(2003)]{GustafssonEE03}
Gustafsson B., Edvardsson B., Eriksson K., et al. 2003, in Stellar 
Atmosphere Modeling, ed. I. Hubeny, D. Mihalas, \& K. Werner, ASP Conf. Ser., 288, 331 

\bibitem[Hill et al.(2002)]{HillPC02} 
Hill, V., Plez, B., Cayrel, R., et al.\ 2002, \aap, 387, 560. doi:10.1051/0004-6361:20020434

\bibitem[Holmberg et al.(2009)]{HolmbergNA09} 
Holmberg, J., Nordstr{\"o}m, B., \& Andersen, J.\ 2009, \aap, 501, 941. doi:10.1051/0004-6361/200811191

\bibitem[Honda et al.(2006)]{HondaAI06} 
Honda, S., Aoki, W., Ishimaru, Y., et al.\ 2006, \apj, 643, 1180. doi:10.1086/503195

\bibitem[Honda et al.(2007)]{HondaAI07} 
Honda, S., Aoki, W., Ishimaru, Y., et al.\ 2007, \apj, 666, 1189. doi:10.1086/520034

\bibitem[James et al.(2004a)]{JamesFB04a} 
James, G., Fran{\c{c}}ois, P., Bonifacio, P., et al.\ 2004a, \aap, 414, 1071. doi:10.1051/0004-6361:20034014

\bibitem[James et al.(2004b)]{JamesFB04b} 
James, G., Fran{\c{c}}ois, P., Bonifacio, P., et al.\ 2004b, \aap, 427, 825. doi:10.1051/0004-6361:20041512

\bibitem[Johnson et al.(2015)]{JohnsonRP15} 
Johnson, C.~I., Rich, R.~M., Pilachowski, C.~A., et al.\ 2015, \aj, 150, 63. doi:10.1088/0004-6256/150/2/63

\bibitem[Lallement et al.(2019)]{LallementBV18} 
Lallement, R., Babusiaux, C., Vergely, J.~L., et al.\ 2019, \aap, 625, A135

\bibitem[Lodders et al.(2009)]{LoddersPG09} 
Lodders, K., Palme, H., \& Gail, H.-P.\ 2009, Landolt B\&ouml;rnstein, 4B, 712. doi:10.1007/978-3-540-88055-4\_34

\bibitem[Marigo et al.(2017)]{MarigoGB17} 
Marigo, P., Girardi, L., Bressan, A., et al.\ 2017, \apj, 835, 77

\bibitem[Marino et al.(2015)]{MarinoMK15} 
Marino, A.~F., Milone, A.~P., Karakas, A.~I., et al.\ 2015, \mnras, 450, 815. doi:10.1093/mnras/stv420

\bibitem[Mucciarelli et al.(2023)]{MucciarelliMB23} 
Mucciarelli, A., Minelli, A., Bellazzini, M., et al.\ 2023, \aap, 671, A124. doi:10.1051/0004-6361/202245133

\bibitem[Nordstr{\"o}m et al.(2004)]{NordstromMA04} 
Nordstr{\"o}m, B., Mayor, M., Andersen, J., et al.\ 2004, \aap, 418, 989. doi:10.1051/0004-6361:20035959

\bibitem[Pace(2013)]{Pace13} 
Pace, G.\ 2013, \aap, 551, L8. doi:10.1051/0004-6361/201220364

\bibitem[Peterson \& Schrijver(1997)]{PetersonS97} 
Peterson, R.~C. \& Schrijver, C.~J.\ 1997, \apjl, 480, L47. doi:10.1086/310607

\bibitem[Peterson(2011)]{Peterson11} 
Peterson, R.~C.\ 2011, \apj, 742, 21. doi:10.1088/0004-637X/742/1/21

\bibitem[Peterson(2013)]{Peterson13} 
Peterson, R.~C.\ 2013, \apjl, 768, L13. doi:10.1088/2041-8205/768/1/L13

\bibitem[Plez(2008)]{Plez08} 
Plez, B.\ 2008, Physica Scripta Volume T, 133, 014003. doi:10.1088/0031-8949/2008/T133/014003

\bibitem[Plez(2012)]{Plez-code12} 
Plez, B.\ 2012, Turbospectrum: Code for spectral synthesis, ascl:1205.004

\bibitem[Prantzos et al.(2020)]{PrantzosAC20} 
Prantzos, N., Abia, C., Cristallo, S., et al.\ 2020, \mnras, 491, 1832. doi:10.1093/mnras/stz3154

\bibitem[Roederer et al.(2010)]{RoedererCK10} 
Roederer, I.~U., Cowan, J.~J., Karakas, A.~I., et al.\ 2010, \apj, 724, 975. doi:10.1088/0004-637X/724/2/975

\bibitem[Roederer et al.(2011)]{RoedererMS11} 
Roederer, I.~U., Marino, A.~F., \& Sneden, C.\ 2011, \apj, 742, 37. doi:10.1088/0004-637X/742/1/37

\bibitem[Roederer \& Sneden(2011)]{RoedererSneden11} 
Roederer, I.~U. \& Sneden, C.\ 2011, \aj, 142, 22. doi:10.1088/0004-6256/142/1/22

\bibitem[Roederer \& Thompson(2015)]{RoedererThompson15} 
Roederer, I.~U. \& Thompson, I.~B.\ 2015, \mnras, 449, 3889. doi:10.1093/mnras/stv546

\bibitem[Roederer et al.(2016)]{RoedererMB16} 
Roederer, I.~U., Mateo, M., Bailey, J.~I., et al.\ 2016, \mnras, 455, 2417. doi:10.1093/mnras/stv2462

\bibitem[Ryan et al.(2001)]{RyanBK01} 
Ryan, S.~G., Beers, T.~C., Kajino, T., et al.\ 2001, \apj, 547, 231. doi:10.1086/318338

\bibitem[Siqueira Mello et al.(2014)]{SiqueiraHB14} 
Siqueira Mello, C., Hill, V., Barbuy, B., et al.\ 2014, \aap, 565, A93. doi:10.1051/0004-6361/201423826

\bibitem[Siqueira Mello et al.(2012)]{SiqueiraBS12} 
Siqueira Mello, C., Barbuy, B., Spite, M., et al.\ 2012, \aap, 548, A42. doi:10.1051/0004-6361/201220100

\bibitem[Siqueira Mello et al.(2013)]{SiqueiraSB13} 
Siqueira Mello, C., Spite, M., Barbuy, B., et al.\ 2013, \aap, 550, A122. doi:10.1051/0004-6361/201219949

\bibitem[Siqueira-Mello et al.(2015)]{SiqueiraAB15} 
Siqueira-Mello, C., Andrievsky, S.~M., Barbuy, B., et al.\ 2015, \aap, 584, A86. doi:10.1051/0004-6361/201526695

\bibitem[Sk{\'u}lad{\'o}ttir et al.(2020)]{SkuladottirHC20} 
Sk{\'u}lad{\'o}ttir, {\'A}., Hansen, C.~J., Choplin, A., et al.\ 2020, \aap, 634, A84. doi:10.1051/0004-6361/201937075

\bibitem[Sneden et al.(1996)]{SnedenMP96}
Sneden, C., McWilliam, A., Preston, G.~W., et al.\ 1996, \apj, 467, 819. doi:10.1086/177656

\bibitem[Sneden et al.(2003)]{SnedenCL03} 
Sneden, C., Cowan, J.~J., Lawler, J.~E., et al.\ 2003, \apj, 591, 936. doi:10.1086/375491

\bibitem[Sneden et al.(2008)]{SnedenCG08} 
Sneden, C., Cowan, J.~J., \& Gallino, R.\ 2008, \araa, 46, 241. doi:10.1146/annurev.astro.46.060407.145207

\bibitem[Sobeck et al.(2011)]{SobeckKS11} 
Sobeck, J.~S., Kraft, R.~P., Sneden, C., et al.\ 2011, \aj, 141, 175. doi:10.1088/0004-6256/141/6/175

\bibitem[Spite et al.(2017)]{SpitePG17} 
Spite, M., Peterson, R.~C., Gallagher, A.~J., et al.\ 2017, \aap, 600, A26. doi:10.1051/0004-6361/201630058

\bibitem[Spite et al.(2018)]{SpiteSB18} 
Spite, F., Spite, M., Barbuy, B., et al.\ 2018, \aap, 611, A30. doi:10.1051/0004-6361/201732096

\bibitem[Spite et al.(2022)]{SpiteSC22} 
Spite, M., Spite, F., Caffau, E., et al.\ 2022, \aap, 667, A139. doi:10.1051/0004-6361/202243960 (Paper I)

\bibitem[Stromgren(1987)]{Stromgren87} 
Str\"omgren, B.\ 1987, The Galaxy, 207, 229

\bibitem[Wanajo(2007)]{Wanajo07}
 Wanajo, S.\ 2007, \apjl, 666, L77. doi:10.1086/521724

\bibitem[Worley et al.(2013)]{WorleyHS13} 
Worley, C.~C., Hill, V., Sobeck, J., et al.\ 2013, \aap, 553, A47. doi:10.1051/0004-6361/201321097

\bibitem[Yong et al.(2014a)]{YongABDC14} Yong, D., Alves Brito, A., Da Costa, G.~S., et al.\ 2014a, \mnras, 439, 2638. doi:10.1093/mnras/stu118

\bibitem[Yong et al.(2014b)]{YongRG14} 
Yong, D., Roederer, I.~U., Grundahl, F., et al.\ 2014b, \mnras, 441, 3396. doi:10.1093/mnras/stu806

\end{thebibliography}
\end{document}